\documentclass[ preprint, amsmath, amssymb,
                aps, superscriptaddress, mathtools,
                hyperref, ]{revtex4-1}
\pdfoutput=1

\usepackage{graphicx}
\usepackage{dcolumn}
\usepackage{bm}
\usepackage[linkcolor=blue]{hyperref}

\usepackage{amsmath}
\usepackage{amssymb}
\usepackage{amsfonts}
\usepackage{bm}
\newcommand*\diff{\mathop{}\!\mathrm{d}}

\usepackage{xspace}
\newcommand*{\y}{\ensuremath{\bm{y}}\xspace}
\newcommand*{\x}{\ensuremath{\bm{x}}\xspace}
\newcommand*{\xk}{\ensuremath{\bm{x}_k}\xspace}
\newcommand*{\xj}{\ensuremath{\bm{x}_j}\xspace}
\newcommand*{\ys}{\ensuremath{\bm{y}}s\xspace}
\newcommand*{\xs}{\ensuremath{\bm{x}}s\xspace}
\newcommand*{\xks}{\ensuremath{\bm{x}_k}s\xspace}
\newcommand*{\xset}{\ensuremath{\{\bm{x}\}}\xspace}

\begin{document}
\title{Inferring the shape of data: A probabilistic framework for analyzing experiments in the natural sciences}

\author{Korak Kumar Ray}
\thanks{These authors contributed equally to this work.}
\affiliation{Department of Chemistry, Columbia University, New York, NY 10027 USA}

\author{Anjali R. Verma}
\thanks{These authors contributed equally to this work.}
\affiliation{Department of Chemistry, Columbia University, New York, NY 10027 USA}

\author{\\Ruben L. Gonzalez, Jr.}
\thanks{To whom correspondence may be addressed: Ruben L. Gonzalez, Jr, Department of Chemistry, Columbia University, 3000 Broadway, New York, NY 10027, Tel: (212) 854-1096, Email: rlg2118@columbia.edu; Colin D. Kinz-Thompson, Department of Chemistry, Rutgers University-Newark, 73 Warren St, Newark, NJ 07102, Tel: (973) 353-0671, Email: colin.kinzthompson@rutgers.edu.}
\affiliation{Department of Chemistry, Columbia University, New York, NY 10027 USA}

\author{Colin D. Kinz-Thompson}
\thanks{To whom correspondence may be addressed: Ruben L. Gonzalez, Jr, Department of Chemistry, Columbia University, 3000 Broadway, New York, NY 10027, Tel: (212) 854-1096, Email: rlg2118@columbia.edu; Colin D. Kinz-Thompson, Department of Chemistry, Rutgers University-Newark, 73 Warren St, Newark, NJ 07102, Tel: (973) 353-0671, Email: colin.kinzthompson@rutgers.edu.}
\affiliation{Department of Chemistry, Rutgers University-Newark, Newark, NJ 07102 USA}

\maketitle

\section*{ABSTRACT}
\noindent 
A critical step in data analysis for many different types of experiments is the identification of features with theoretically defined shapes in N-dimensional datasets; examples of this process include finding peaks in multi-dimensional molecular spectra or emitters in fluorescence microscopy images. Identifying such features involves determining if the overall shape of the data is consistent with an expected shape, however, it is generally unclear how to quantitatively make this determination. In practice, many analysis methods employ subjective, heuristic approaches, which complicates the validation of any ensuing results—especially as the amount and dimensionality of the data increase. Here, we present a probabilistic solution to this problem by using Bayes’ rule to calculate the probability that the data has any one of several potential shapes. This probabilistic approach may be used to objectively compare how well different theories describe a dataset, identify changes between datasets, and detect features within data using a corollary method called Bayesian Inference-based Template Search (BITS); several proof-of-principle examples are provided. Altogether, this mathematical framework serves as an automated `engine' capable of computationally executing analysis decisions currently made by visual inspection across the sciences.

\frenchspacing 

\section{Introduction}
\begin{figure}
\centering
\includegraphics[width=\columnwidth]{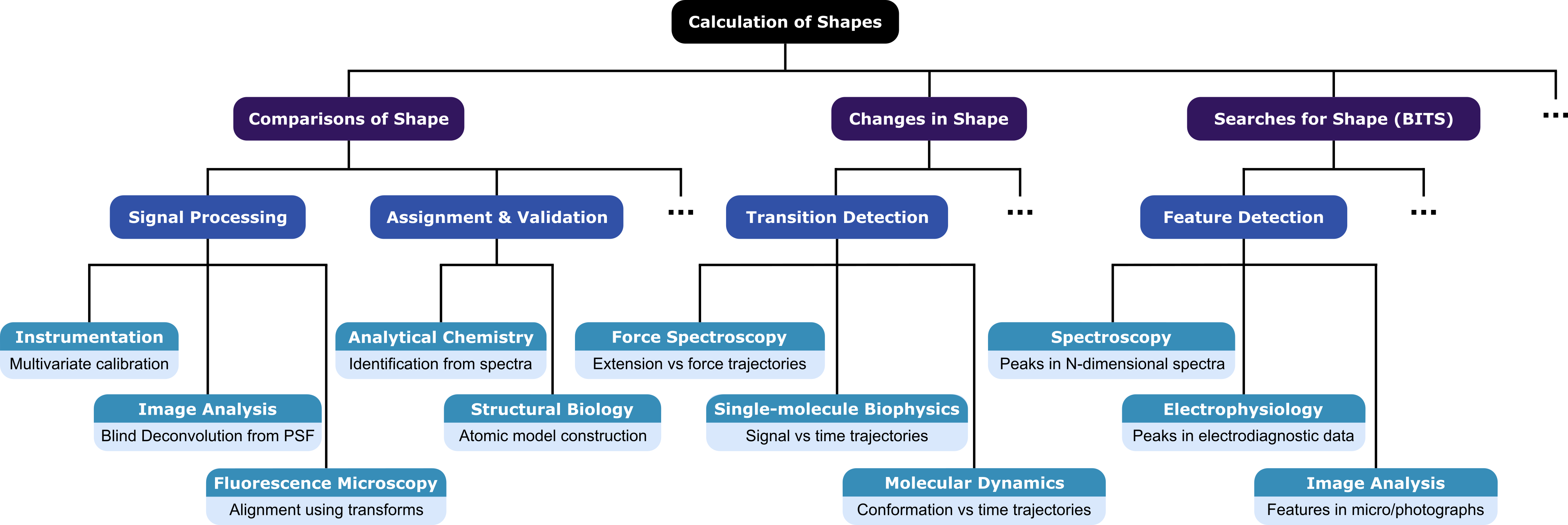}
\caption{\textbf{Applications of shape calculation to the physical and life sciences.} A graphical representation of our mathematical framework (black), with examples of data analysis methods made possible by it (purple), specific tasks these methods enable (blue), and applications of these tasks in specific techniques in the physical and life sciences (dark cyan) along with the specific problems (light cyan) that the application of our framework to these techniques may address. The given examples are not meant to be exhaustive.} \label{fig1}
\end{figure}

\noindent
Across the physical and life sciences, many experimental techniques rely upon pragmatic data analysis steps where an expert researcher is required to make scientific decisions based on their visual perception of data. This perception involves identifying and recognizing correlations between datapoints that stem from underlying physical processes, which are ideally invariant across experiments; we refer to these correlations as the latent structure of the data. Latent structure manifests visually in what we would colloquially call the `shape' of the data and is the basis for inspection-driven analysis decisions. For example, an expert researcher might have to visually identify a feature of interest by recognising an expected shape in a plot of their data (\textit{e.g.}, a shoulder on a peak in a molecular spectrum). Alternatively, such a researcher might anticipate the location of an expected feature within their plotted data (\textit{e.g.}, a peak at a specific frequency in a molecular spectrum), but must then decide whether or not it is actually present at that location. In these types of determinations, the researcher must generate at least two visual models of a phenomenon, manually compare those models to the shape of their experimental data, and then choose the model that, in their expert opinion, best describes the data. To be explicit, in the first example above, the researcher visually compares both the shape of a peak and the shape of a peak with a shoulder to the experimental molecular spectrum. Similarly, in the second example, the researcher visually compares both the shape of a peak and the shape of signal-free background noise 
to the experimental molecular spectrum.
\\
\indent \indent 
A key advantage of such expert-driven analyses is the human ability to make accurate, informed decisions about the latent structure of experimental data, even in the absence of a full theoretical description of the phenomenon of interest. For instance, while the spectral line shapes of peaks in molecular spectra arise due to physical processes with well established theoretical foundations, a full quantum mechanical calculation is generally not required to determine whether a certain peak exists at a particular location, nor whether or not it has a shoulder. Instead, approximate models of the shape of a peak, guided by a researcher's physics-based intuition and years of experience, are usually sufficient for the level of analysis required for these problems. Having considered all the models they deem appropriate, the expert researcher then decides which of those models is the best description of their data and, thus, is best supported by the available evidence. 
\\
\indent \indent 
Such researcher-dependent approaches to data analysis create major practical, quantitative, and scientific challenges. An obvious difficulty is the time required for manual data processing, which limits a researcher’s output and productivity. Another is simply the learning curve required to perform visual inspection-based analysis tasks--an extensive amount of training is required before an inexperienced researcher can build enough physics-guided intuition to accurately and reliably interpret experimental data. Yet another obstacle is the lack of a quantitative metric for assessing the confidence one should have in one’s own or someone else’s analysis decisions, especially in cases of conflicting results. The lack of such a quantitative confidence metric makes it similarly difficult to validate or replicate such visual inspection-based analyses. Most importantly, there exist fundamental barriers which inhibit precise communication of the details of these analyses: namely, the intrinsic complexity of describing in writing the exact details of a method performed within one’s mind, and conversely, of understanding the details of such a method solely by reading a description of it. All of these challenges are exacerbated as scientific research fields progress towards more quantitative, data-driven approaches, and as more techniques are developed that yield larger amounts of increasingly more complex data, as is systematically occurring with, for instance, the advent of ultrahigh-throughput methods \cite{Johnstone2009, Leek2010}.
\\
\indent \indent 
In contrast to such human-dependent approaches, here we have developed a computational framework designed to automate and imitate the visual inspection-based data analysis steps typically performed by expert researchers, but in manner that is quantifiable, reproducible, and precisely communicable. Inspired by the human ability to visually assess the `shape' of plotted data, our approach is to use probabilistic inference-based model selection \cite{Jaynes2003, Bishop2006} on technique-specific sets of models in order to calculate how well the shape of each model, which we call a `template', can quantitatively describe the latent structure of the data. Specifically, we apply Bayes' rule to the probability expressions, known as evidences, which here characterise the degree to which the models under consideration can explain the observed data, in a process known as Bayesian model selection (BMS) \cite{CKT2021}. Broadly, the advantages of adopting a Bayesian framework have led to the increased usage of Bayesian methods in recent times across the sciences \cite{malakoff1999}. For instance, in the field of biophysics, and particularly single-molecule studies, the use of Bayesian inference has been transformative due to its intrinsic ability to handle particularly noisy data (reviewed in \cite{CKT2021}). However, the difficulty of deriving evidences has historically limited the extension of Bayesian inference to BMS-based analysis approaches \cite{Toussaint2011, Friel2012}, except in a few specialised cases (\textit{e.g.}, in \cite{Cossio2013} and \cite{Ensign2010}). 
\\
\indent \indent 
In this work we create a generalised BMS-based framework using closed-form expressions for evidences that can be adapted by researchers in the physical and life sciences to a variety of different applications with computational ease and efficiency (Fig. \ref{fig1}). Additionally, because each implementation of this framework is defined by the specific set of physics-informed models considered, our approach can be leveraged to create constrained analyses that achieve optimal balances between theoretical precision and computational efficiency. We also harness this framework to create a corollary method, called Bayesian Inference-based Template Search (BITS), that enables us to achieve a large computational speedup when identifying and localising multiple features of interest within a dataset. Altogether, our probabilistic, BMS-based framework is a radically new method for analysing data that allows researchers to computationally mimic expert-based visual analyses without needing to resort to a subjective, researcher-dependent approach.

\section{Describing the Shape of a Dataset} \label{sec:math}
\noindent
In this section, we detail a mathematical framework designed to resemble the process of expert-based visual analysis. This approach uses orientation-preserving affine transformations of a template vector to map the associated model of latent structure onto the experimental data being analysed. The marginal likelihood of the data given a specific template is then calculated regardless of the scaling and translation of this transformation, or the noise present in the data. These marginal likelihoods are computed for a predefined set of templates, and are then used in BMS to calculate the posterior probabilities for each template. The shape of the data is then optimally described by the template with the highest posterior probability.

\subsection{Defining a Template} \label{sec2i}
\noindent
To begin, we consider the problem of specifying a model of the latent structure of a dataset for the purpose of mimicking visual recognition. In our framework, a dataset, \y, is a tensor whose components are the individual, scalar datapoints. Regardless of the experimental relationships between those datapoints (\textit{i.e.}, the organisational structure of the tensor), for simplicity we can reshape \y into an N-dimensional vector $\y=[y_1,...,y_N]$, where $y_i$ are the scalar datapoints. One can imagine a dataset \y collected using a particular instrument, in a particular location, on a particular day, and by a particular researcher. Altogether, these specific factors might induce systematic differences in \y relative to an otherwise equivalent experiment. For example, an optical filter in an instrument might slowly oxidise, which could reduce the intensity of light incident upon the detector and, over a period of months, yield \y with different scales (\textit{i.e.}, units). Similarly, overhead lights might be left on accidentally when making an optical measurement, which could increase the background photons incident on the detector and yield \y with different relative offsets. Likewise, local vibrations might vary from day to day, which could affect the stability of an instrument and yield \y with different amounts of noise. Yet factors such as the scaling, offset, or amount of noise in a measurement generally do not alter the underlying physical processes that give rise to the measured data, and thus should not affect the latent structure  (\textit{i.e.}, the shape) of \y. Instead, these factors often act as irregularities, or nuisances, that can limit our ability to model the shape of \y, hence our use of the term `nuisance parameters' to describe them. With this in mind, we define a template, $\x=[x_1,...x_N]$, as a particular N-dimensional vector of scalar quantities which is related to \y through the following transformation \begin{equation}
    y_i=mx_i+b+\xi_i \quad  \text{for all}  \quad i=1,\text{…},N. \label{eqn:affine}
\end{equation}
In this equation, $m$ and $b$ are nuisance parameters representing changes to scale and offset, respectively, $\bm{\xi}=[\xi_1,...,\xi_N]$ is a nuisance parameter composed of stochastic terms representing the experimental noise, and $N$ is the number of components in \x or \y.  If we recall our definition of shape as correlations within data which derive from fundamental physical processes, we can conceptually understand a template, \x, as an ideal representation of these correlations, without noise or background.
To avoid confusion, we note that our definition of shape is distinct from those that take shape to mean a boundary or segmentation in data \cite{Dryden2016}, and that it is this choice of definition which enables our framework and aligns it with the intuitive visual analyses performed manually by researchers.
\\
\indent \indent 
It is important to note that the shape of \y, regardless of any distortions caused by the experimental nuisance parameters described above, may often be reasonably described by many different \xs. Indeed, there are no restrictions on what specific \xs one may choose as templates. Different \xs might depend upon different levels of theory, the particular details of the experimental setup, and even sample-to-sample variability. For example, the laws of diffraction dictate the point spread function (PSF) that describes the shape of point emissions in a fluorescence microscopy image \cite{Pawley2006}. However, for a standard microscope, the PSF can be modeled by an Airy disk; a two-dimensional circular Gaussian function; or, in order to incorporate an astigmatism correction, even a two-dimensional elliptical Gaussian function \cite{Shen2017}. Each of these models of the PSF provides a distinct, theoretically valid \x capable of modeling the shape of \y with varying degrees of complexity. Alternatively, one’s \xs could be empirically derived from data previously recorded in other experiments. In the context of the above example, an \x can be created from fluorescence microscopy images of point emitter-like samples without needing to explicitly invoke a theory of diffraction, and such empirically derived \xs might even model the latent structure of \y more effectively than analytical, theory-derived \xs. Regardless of the complexity of \x or its origin, once formulated, it is directly related to a \y by the simple affine transformation given in Eqn. \eqref{eqn:affine}.
\\
\indent \indent 
The choice of which \x (or set of \xs) one uses to model the shape of \y depends not only on the experimental technique but also on the level of precision required for that particular analysis. When using this mathematical framework to analyse experimental data from the natural sciences, one can invoke prior knowledge of the physics governing the experiment which gave rise to the data to constrain the choice of templates used in the analysis. Thus, while templates with higher complexities (\textit{e.g.}, an Airy disk as a model for a PSF) may be required for certain applications, in other cases, a less complex template (\textit{e.g.}, a 2D Gaussian as model for a PSF) can perform just as effectively while greatly reducing the computational cost of the analysis. The flexibility in choice of templates enabled by our framework can greatly increase the efficiency and effectiveness of an analysis method (see Sec. \ref{BITS}), however, determining which of the chosen \xs, if any, is the optimal template requires that we first compute how well the shape of \y is explained by a given \x.

\subsection{Deriving Probabilistic Expressions for Shape} \label{sec2ii}
\noindent
After defining an \x, we quantify the degree to which it describes the latent structure of \y, regardless of the nuisance parameters described above in Eqn. \eqref{eqn:affine}. For the $k$\textsuperscript{th} template, \xk, in a set of templates, $\xset = \{\bm{x}_1, ..., \bm{x}_K \}$, this means calculating a marginalised likelihood probability called the evidence, $P(\y | \xk, M_0)$. Here, the conditional $M_0$ represents all of the details of the experiment, previous knowledge about the system, and particulars of the analysis method(s)—including which templates have been incorporated into the chosen \xset. The expression for the evidence of \xk is the marginalisation of the joint probability, which is given by
\begin{equation}
    P\left(\y \, \vert \, \xk,M_0 \right) = \iiint p\left(\y \, \vert \, \xk, \bm{\xi},m,b,M_0\right)
    \times p\left(\bm{\xi},m,b \, \vert \, M_0\right) \diff \bm{\xi} \diff m \diff b. \label{eqn:evidence}
\end{equation}
In Eqn. \eqref{eqn:evidence}, $p(\y \vert \xk,\bm{\xi},m,b,M_0)$ is called the likelihood, and it represents the probability density of observing \y for a given \xk and given values of the nuisance parameters; $p(\bm{\xi},m,b \, \vert \, M_0)$ is called the prior, and it represents the joint probability density of those particular nuisance parameter values based on the prior knowledge specified by $M_0$.
\\
\indent \indent 
In this work, we have used combinations of different likelihoods and priors to derive a set of evidences, expressed in closed-form, that are particularly useful for calculating the shape of data in a variety of experimental situations. For all of the cases presented here, we have assumed in our $M_0$ that the $\xi_i$ are uncorrelated, such that $\langle \xi_i \rangle$ = 0, and $\langle \xi_i,\xi_j \rangle$ = $ \tau^{-1} \delta_{ij}$, where $\tau$ is a constant called the precision and $\delta_{ij}$ is the Kronecker delta. While this assumption is not a requirement of our approach, this noise model is often experimentally reasonable, and it has allowed us to present analytical solutions to evidence integrals in many general situations (see Supplemental Materials, Section 2 for other noise models). Together with Eqn. \eqref{eqn:affine}, this assumption yields the following likelihood function:
\begin{equation}
    p(\y\vert\,\xk,m,b,\tau,M_0)= \prod_{i=1}^N \left(\frac{\tau}{2\pi}\right)^{1/2}  e^{-\frac{\tau}{2} \left(y_i-mx_i-b\right)^2}. \label{eqn:likelihood}
\end{equation}
Very similar likelihood functions arise with this noise model when $m$ is known to be 0 or 1, and/or $b$ is 0. 
\\
\indent \indent 
Specifying the probability expression for the prior--the second term in the integrand in Eqn. \eqref{eqn:evidence}--requires that we mathematically represent our previous knowledge of how $m$, $b$, and $\tau$ are distributed in the experiments of interest \cite{Jaynes2003}. In particular, the prior dictates the integration bounds of Eqn. \eqref{eqn:evidence} by determining the values that are possible for these parameters to assume (\textit{i.e.}, regions where the prior probability is non-zero). For the results derived here, we have used so-called ‘maximum entropy’ priors, which allow us to encode information and constraints into our prior probability expressions, without dictating their functional form in an \textit{ad hoc} manner \cite{Jaynes2003}. If we assume in $M_0$ that we only know that $m$, $b$, and $\tau$ are within some range and that we do not know the magnitude of $\tau$ (\textit{i.e.}, the amount of noise we expect), then the corresponding maximum entropy priors are a uniform distribution for $m$ and $b$, and a uniform distribution over the logarithm of $\tau$ (see Supplemental Materials, Section 1). If we further assume that $m$, $b$, and $\tau$ are independent, then the corresponding joint prior of these parameters within the given ranges is
\begin{equation}
    p(m,b,\tau\,|\,M_0)= \frac{\tau^{-1}}{\Delta m \Delta b \Delta (\ln \tau)}, \label{eqn:prior}
\end{equation}
where the shorthand $\Delta f \equiv f_{max}-f_{min}$ defines the range of a parameter.
\\
\indent \indent 
In order to analytically integrate Eqn. \eqref{eqn:evidence}, the integration bounds in the prior must be explicitly defined. We note that a positive transformation of an \xk ($m > 0$) can have a distinct physical interpretation from a negative transformation ($m < 0$). Thus, in order to differentiate between these two cases and properly model the underlying shape of \y, we impose that \xk and \y be oriented in the same direction. This constraint can be encoded into the calculation by only considering orientation-preserving (\textit{i.e.}, positive scaling) affine transforms of the \xk in Eqn. \eqref{eqn:affine}. To explicitly include this information in the prior, and thus in our $M_0$, we therefore use $m_{min}=0$ rather than some $m_{min}<0$. In the case that the negative transformation ($m < 0$) is of interest, we note that $-\xk$ with $m > 0$ is equivalent to $\xk$ with $m < 0$. Closed-form expressions for the evidence derived using other integration bounds are also provided in the Supplemental Materials. Additionally, to keep the prior normalised and avoid using a so-called ‘improper’ prior, the minimum and maximum values must be chosen such that $\Delta m$, $\Delta b$, and $\Delta \ln \tau$ are not infinite. For the purposes of a tractable integration \cite{Gradshteyn2014, Ng1969}, we have used such large negative and positive values that the integration bounds in Eqn. \eqref{eqn:evidence} can be approximated as $m \in [0,\infty)$, $b \in (-\infty,\infty)$, and $\tau \in [0,\infty)$. While the exact values of the bounds are important and should be chosen judiciously, we note that the resulting prior normalisation terms end up canceling in subsequent steps during BMS (see below). Using the integration bounds discussed above, the closed-form probability expression for the evidence calculated using Eqn. \eqref{eqn:evidence} is 

\begin{multline}
P(\y\,\vert\,\xk,M_0) = \frac{\Gamma(\frac{N-2}{2}) N^{-\frac{N}{2}} \pi^{-\frac{N-2}{2}}}{2 \Delta m \Delta b \Delta \ln \tau} V_x^{-\frac{1}{2}} \\
\times \left( V_y \left( 1-r^2 \right) \right)^{-\frac{N-2}{2}} \left[ 1 + \frac{r}{|r|} I_{r^2} \left(\frac{1}{2}, \frac{N-2}{2} \right) \right], \label{eqn:closedevidence}
\end{multline}
where $V_x \equiv \langle \xk^2 \rangle - \langle \xk \rangle^2$, $V_y \equiv \langle \y^2 \rangle - \langle \y \rangle^2$, $r \equiv \frac{\langle \xk\y \rangle - \langle \xk \rangle \langle \y \rangle}{\sqrt{V_x V_y}}$, $ \langle f \rangle \equiv \frac{1}{N} \sum_i^N f_i$  is the arithmetic mean, $\Gamma(x)$ is the gamma function, and $I_x(a,b)$ is the regularised incomplete beta function. This evidence is the probability that the shape of \y corresponds to a specific template \xk, regardless of the particular values of the (positive) scale, offset, and noise parameters used in the affine transformation that relates \xk to \y (Eqn. \eqref{eqn:affine}). At first glance, the appearance of the term $V_x^{-\frac{1}{2}}$ suggests that two \xks that are equivalent up to a multiplicative constant would have different abilities to explain the same \y. However, that constant must also be accounted for in the prior term $\Delta m^{-1}$, where it can cancel this effect. Thus, choosing the range for $m$ in the prior is intimately related to setting the $V_x$ of the \xk and, unless one has a reason to believe different models have different ranges of $m$, the \xk within a $\{x\}$ should be normalised such that their $V_x$ are equivalent.
\\
\indent \indent 
While the evidence expression in Eqn. \eqref{eqn:closedevidence} is very general in the sense that it can be used for almost all choices of templates, it is not applicable to the special, `null' case in which a template is absent (\textit{i.e.}, where \xk is flat and/or $m$ is only zero).  This case is very useful in our approach for validating the presence or absence of a shape in experimental data, as we will show in following section.  The corresponding evidence expression for this case is 
\begin{equation}
    P(\y\,\vert\, \x_{null},M_0) = \frac{\Gamma(\frac{N-1}{2}) N^{-\frac{N}{2}}}{\Delta b \Delta \ln \tau} (\pi V_y )^{-\frac{N-1}{2}}, \label{eqn:null}
\end{equation}
where $\x_{null}$ represents the case that the model lacks a template. This evidence expression represents the probability that the experimental data is featureless (\textit{i.e.}, lacking any latent structure) beyond the presence of a constant background offset and noise, regardless of the exact values of these parameters. Together, the evidence expressions in Eqns. \eqref{eqn:closedevidence} and \eqref{eqn:null} enable us to quantitatively express how well the shape of experimental data agrees with candidate templates, independent of extraneous details and nuisance parameters that may change from experiment to experiment. 

\subsection{Describing the Shape of a Dataset using Bayesian Model Selection (BMS)} \label{sec2iii}
\noindent
We compare the performance of different templates by using BMS \cite{Jaynes2003, Bishop2006, CKT2021} in order to calculate the probability that each \xk is the best description of the shape of the data, \y. This calculation is conditionally dependent on the assumptions in $M_0$, which define the specifics of the analysis method, including the composition of \xset. Multiple distinct analysis methods can consequently be developed by using different $M_0$s to tailor their effectiveness to individual experimental situations and systems. For any chosen $M_0$, an appropriate template prior probability for \xk, $P(\xk |M_0)$, must then be assigned, for example, by: (i) using an equal \textit{a priori} assignment of $K^{-1}$, where $K$ represents the number of templates in \xset; (ii) learning prior values from separate experiments; or even (iii) using a Dirichlet process or hierarchical Dirichlet process \cite{Teh2006} for a non-parametric ‘infinite’ set of templates. Once all of the $P(\xk |M_0)$ have been assigned, Bayes’ rule can be used to perform BMS and compute the template posterior probability as
\begin{equation}
    P(\xk \,\vert\, \y,M_0) = \frac{P(\y \,\vert\, \xk,M_0) P(\xk \,\vert\, M_0)}{\sum_{j=1}^K P(\y \,\vert\, \xj, M_0) P(\xj \,\vert\, M_0)}. \label{eqn:bayes}
\end{equation}
This expression represents the probability of an \xk given the observed data \y and, thus, may be used to identify the \xk in \xset that most optimally describes the latent structure of \y (for a specific choice of $M_0$). Using Eqn. \eqref{eqn:bayes} is therefore a quantitative means by which the underlying shape of experimental data may be determined. Furthermore, by considering a `background'-shaped \xk and/or just the presence of noise (\textit{i.e.}, Eqn. \eqref{eqn:null}) in the BMS process, this approach can also validate whether using the most probable \xk to describe the shape of \y is justified, or whether the shape of \y can be better explained as just noise in the data. Altogether, this BMS process sets up an objective, quantitative, researcher-independent metric for not only determining the shape of experimental data, but also validating such shape assignments.
\\
\indent \indent 
The shape-calculation equations we report above describe a relationship between ideal distributions (\textit{i.e.}, \xk) and noisy signals (\textit{i.e.}, \y) that is independent of many experimental details which would otherwise complicate the analysis being performed. The only requirements are that both \xk and \y exist in the same data-space and are vectors of the same size. Practically, however, most templates are generated from some underlying model that exists in a separate ‘model-space’ distinct from the data-space of \xk and \y. Relating such a model-space to data-space requires that a set of parameters, $\{ \theta \}$, be used to map the model to an \xk. For example, a model of a three-dimensional object being projected onto a two-dimensional image may use the Euler angles of the object to generate \xks with different orientations in the two-dimensional image data-space.
\\
\indent \indent 
Generally, when using such a model to generate \xks for identifying the shape of \y, the template posterior probabilities for an entire group of model-associated \xks must be calculated to account for the many possible ways that the single model could have been mapped into data-space. Having performed all of those calculations, it is then possible to marginalise out the dependence upon some of the $\{ \theta \}$ from the model. In the example above, marginalising out the Euler angles would yield the posterior probability that the shape of \y corresponds to a two-dimensional image of the model, regardless of not knowing the true orientation of the three-dimensional object being projected into the image. Thus, this type of marginalisation in data-space enables our framework to provide objective measures for shape assignment in model-spaces as well. We note that the map between model-space and data-space used in these shape-calculations should be explicitly acknowledged and defined in order to mitigate unintentional mis-estimations of the weight of particular models in data-space during the change of variables. Finally, it is worth mentioning that such model-spaces almost always exist for scientific analyses, even if they are only implicitly invoked within $M_0$. 
\\
\indent \indent 
The most complete implementations of these BMS-based shape calculations occur when using $M_0$s that specify every physically appropriate \xk. However, this approach may not always be theoretically possible nor computationally feasible if an effectively infinite number of \xks exist. In such situations, it is worth noting that, depending on the precision required by a particular analysis method, the full set of templates may not be required to obtain effective results. Importantly, a major benefit of BMS is that we can determine which \xk among a set of approximate templates best describes the shape of \y, even if none are `exact'. Additionally, the BMS expression in Eqn. \eqref{eqn:bayes} can be rearranged into a function of the log difference of evidence expressions (\textit{i.e.}, a Bayes’ factor) between a test \xk and an appropriate control \xk (or a ‘null’ model), which yields an effective cost function for the direct optimisation of a single \xk (see Supplemental Materials, Section 3). Overall, the most powerful aspect of the BMS-based shape calculations described here is that by considering different $M_0$s, an analysis method can be optimised for completeness (where all appropriate templates are enumerated) or efficiency (where only a test and a control template are considered), or for a trade-off between the two (using only a restricted set of templates), as the situation demands. This flexibility is a large reason why our framework can be effectively adapted to mimic nearly any of the subjective, expert-based analysis methods that it is meant to replace. Furthermore, the ability to easily disseminate the \xset used in an analysis means that methods can be readily shared, critiqued, and reproduced. Together, these especially powerful aspects of our framework make it extremely straightforward to implement tailor-made, shape-based analysis methods for new experiments and applications.

\section{Searching for Shapes: Bayesian Inference-based Template Search (BITS)} \label{BITS}
\noindent
While the practical scientific applications of shape calculations are numerous (see Fig. \ref{fig1} and Sec. \ref{examples}), the flexibility of our framework leads to a corollary of this approach that can be used to search for the presence of particular 'local' features in the data. Experimental examples of this kind of analysis include finding the location of peaks in molecular spectra, puncta in fluorescence microscopy images, or stepping behavior in time-series. In all of these situations, an underlying physical relationship exists between the datapoints in \y (\textit{e.g.}, emission wavelength, Cartesian position on a substrate, or measurement time). In the previous section we considered \y as an $N$-dimensional vector in a manner that largely ignored the relationships between datapoints. Because \y is a tensor, however, we can reshape it to fundamentally account for these relationships. For instance, if \y is a fluorescence microscopy image, then each datapoint might correspond to a pixel of spectral colour $c$ with an associated position ($r_x$, $r_y$, $r_z$) in the sample-space of the experiment. Thus, it would be useful to reshape \y from a first-order tensor (\textit{i.e.}, a vector) of $N$ scalar datapoints into a more natural representation as a fourth-order tensor with one dimension each for $c$, $r_x$, $r_y$, and $r_z$. Because $r_x$, $r_y$, and $r_z$ exist in a Euclidean metric space, we can also calculate a distance, $d(y_i,y_j)$, between any two datapoints in this example. With such a distance metric for at least one of the tensor dimensions of \y, a local neighbourhood around the position of $y_i$ can be defined as the subset of datapoints, $\bm{y}_{(i,\varepsilon)} \subset \y$, for which $d(y_i,y_j) < \varepsilon$, where $\varepsilon$ is a specified distance cutoff. Notably, this neighbourhood contains $n\le N$ datapoints, and may be orders of magnitude smaller depending on the choice of $\varepsilon$. Thus, for a fixed value of $\varepsilon$, \y can be thought of as a composite of approximately $N$ unique $\varepsilon$-neighbourhoods of size $n$ (\textit{i.e.}, distinct subsets $\bm{y}_{(i,\varepsilon)}$).
\\
\indent \indent 
We can then define the local, latent structure of \y by performing the BMS-based shape assignment described in Sec. \ref{sec:math} separately within each of these unique neighbourhoods using a set of templates, \xset, where each \xk is also of size $n$. Intuitively, this process can be understood as `scanning' a small region through \y along the dimensions of the tensor and assessing the shape of the data at each site. Whenever one of the \xk is found to be an appropriate description for the data in a particular neighbourhood, a feature (\textit{i.e.}, \xk) is effectively `localised' at that site. Therefore, we call this local analysis approach `Bayesian inference-based template search' (BITS), because it localises the templates in \xset within \y by traversing the unique neighbourhoods of \y and determining the latent structure of the data in each using the BMS approach described above. As the name BITS suggests, this approach is conceptually similar to traditional template matching calculations (\textit{e.g.}, \textit{via} normalised cross correlation \cite{Briechle2001}), and in fact incredible mathematical similarities, as seen in the $r$ cross-correlation term between \xk and \y in Eqn. \eqref{eqn:closedevidence}, have naturally arisen from our probabilistic approach. As such, we believe many strategies used for template matching (\textit{e.g.}, fast Fourier transforms) might be adapted with future work. Regardless, as discussed below, by casting the template matching process into a probabilistic framework BITS enables powerful extensions facilitated by model selection, such as model comparison and automatic feature localisation.
\\
\indent \indent 
We note that each local calculation is technically performed over all \y but, by splitting the likelihood into two regions, one within $\bm{y}_{(i,\varepsilon)}$ modeled by \xset and one without $\bm{y}_{(i,\varepsilon)}$ modeled by $\x_{null}$ (\textit{i.e.}, non-local data is `noise'), the evidence contribution from without the local region is the same for each \xk and cancels in the Bayes' factors of Eqn. \eqref{eqn:bayes}. Thus, the entire calculation can be simplified, and only the local region within $\bm{y}_{(i,\varepsilon)}$ needs to be addressed. Of course, rather than use the local BITS approach described here, a composite template simultaneously containing all of the features being localised could be used to describe the shape of the entire \y, however, as we will show, BITS is much more computationally efficient. For instance, a \y of size $N$ that contains $R$ unique features of size $n$ that are to be localised with datapoint-resolution would require $N^R$ distinct templates be tested. Both constructing each template and calculating the evidence for a template are $\mathcal{O}(N)$, so such a full-sised shape calculation has computational scaling of $\mathcal{O}(N^{R+1})$. Clearly, this approach has severe scaling issues for any number of features. Fortunately, the equivalent BITS calculation that interrogates $N$ localisation sites, and where we have chosen $\varepsilon$ so that the \xk are the size of the features, $n$,  has a computational scaling of $\mathcal{O}(nNR)$. In the context of shape-based analyses, template searching with BITS greatly reduces the computational burden of localising features down from a geometric to a linear scaling.

\begin{figure}
\centering
\includegraphics[width=0.65\columnwidth]{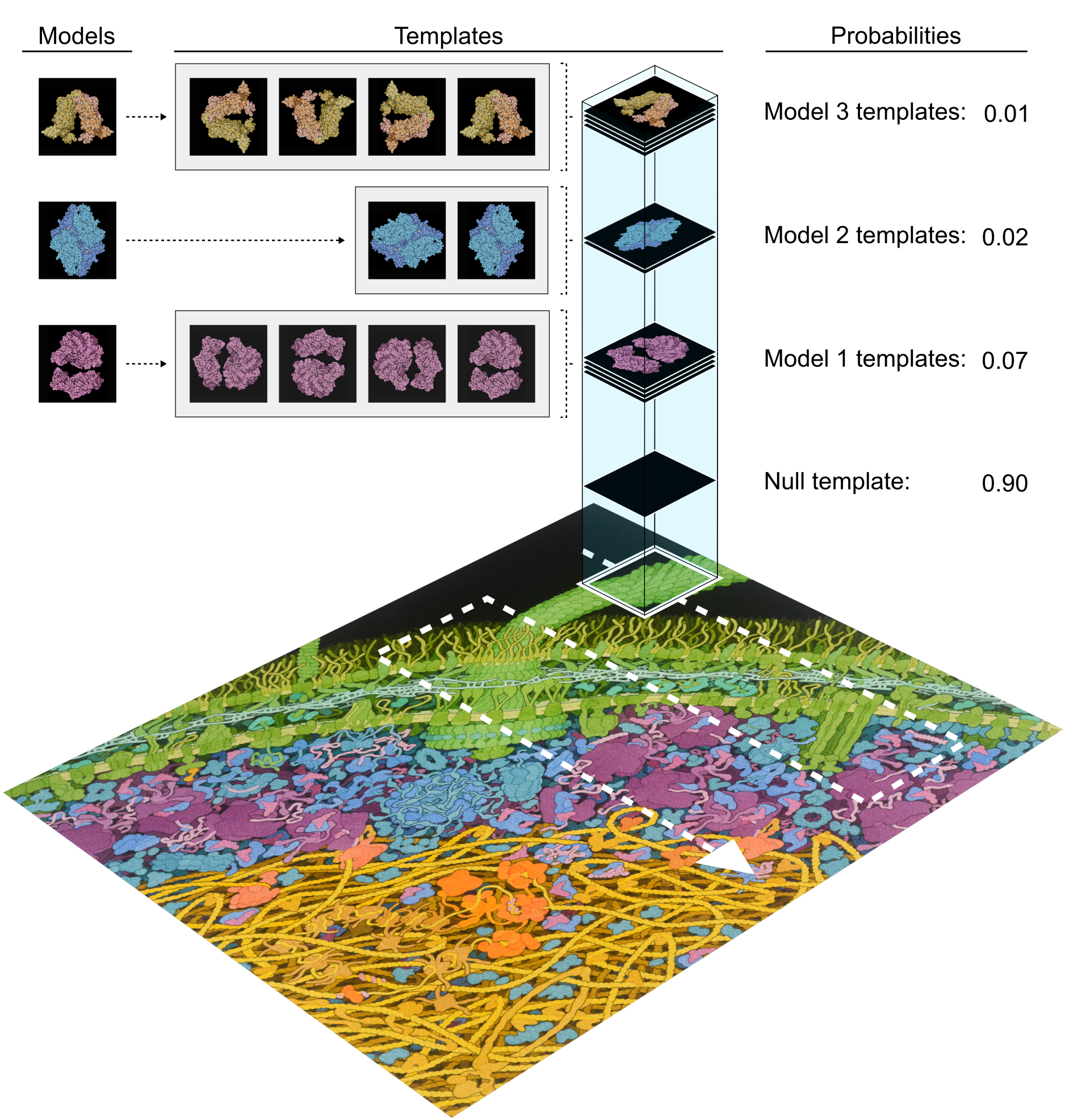}
\setlength{\belowcaptionskip}{-12pt}
\caption{\textbf{Illustration of the Bayesian Inference-based Template Search Algorithm.} An example of a BITS process is shown, where three different biomolecules are searched for in a two dimensional image of a cellular environment. Different sets of cellular components are coloured differently for illustrative purposes, to demonstrate the expected locations of the different biomolecules (green for cell membrane components, purple for translation machinery, blue for enzymes, and yellow and orange for transcription and replication machinery). A set of rotational templates (in grey boxes) generated from different models of the biomolecules of interest (coloured as above) is scanned through subsections of the image (white arrow), and the probability that each template best matches the local shape of the data in a specific subsection (white box) is calculated and then marginalised into an aggregate probability for each model that is used to identify the local composition of the image. The probability values shown were chosen to illustrate the example case of the null template being identified in the case where the shape of a subsection cannot be explained by any of the model templates. Adapted from illustrations by David S. Goodsell, RCSB Protein Data Bank (DOIs: 10.2210/rcsb\_pdb/goodsell-gallery-028, 10.2210/rcsb\_pdb/mom\_2000\_10, and 10.2210/rcsb\_pdb/mom\_2016\_6, 10.2210/rcsb\_pdb/mom\_2003\_3).}
\label{fig2} 
\end{figure}

\indent \indent 
The BITS process is demonstrated in Fig. \ref{fig2} with an illustrative example of the analysis of an image of a cellular environment. The data-space to be analysed in this example consists of a second-order tensor of pixel intensities where the two tensor dimensions correspond to Cartesian coordinates in the cellular environment. While the image is coloured to differentiate and visualise different cellular components with the human eye, we note that, for simplicity, our illustrative example is dealing with the total intensity value of each pixel. Three-dimensional atomic-resolution structural models for these molecules are used to generate a corresponding set of two-dimensional \xks that represent each biomolecule in a particular orientation in the image of the cellular environment (shown in the figure in grey boxes). Given the number of templates, and that the size of these templates is much smaller than the total size of the image, BITS can be used very efficiently in this analysis.
\\
\indent \indent 
Along with a null template ($\x_{null}$), the biomolecular model templates are `scanned' through the image \y, and the BMS calculation of Eqn. \eqref{eqn:bayes} is performed on the $\bm{y}_{(i,\varepsilon)}$ at each site. The white square on the image shows the specific local neighbourhood $\bm{y}_{(i,\varepsilon)}$ currently being interrogated using BMS, and in subsequent steps BMS calculations are performed on the adjacent local neighbourhoods (`scanning' order denoted by the white arrow). The biomolecular orientation dependence of the \xks is marginalised out of this calculation by combining the template posterior probabilities of the \xks derived from the same biomolecular model. This yields the model posterior probability that each biomolecule of interest is localised at a particular position in the image regardless of its orientation. The specific neighbourhood being analysed in Fig. \ref{fig2} highlights the advantages of including a $\x_{null}$ in a BITS analysis. While this region of the image contains some latent structure, we can visually see that it is not  explained by any of the biomolecular model templates. Corresponding to this visual analysis, BITS finds that the null template has the highest posterior probability, and thus, no feature is localised. This stipulation, that a feature is only localised if the shape of the data is better explained by a model template than random noise, provides critical protection against over-fitting, which is a distinct advantage of BITS over other template searching methods. 
\\
\indent \indent 
Notably, in Fig. \ref{fig2}, only a small sample of biomolecular model orientations are used, and we assume that if a particular neighbourhood contains a biomolecule of interest in a similar orientation as one of the chosen templates, there is sufficient overlap of shape between the two for it to be detected in the BMS calculation. The exact amount of similarity required for this correspondence depends only upon how well the other templates can account for the local shape of this neighbourhood. Thus, while the number of sampled orientations may need to be optimised for different applications depending on the desired outcome, \xset with sparse samples of a model's orientations may still be used to effectively localise features.
\\
\indent \indent 
Overall, the BITS algorithm can be understood as a method that quantifies the degree to which the latent structure in the data of each neighbourhood $\y_{i,r}$ of \y is correlated with an \xk. This correlation is maximised when BITS reaches the `true' location of a feature of interest in the data and the \xk is perfectly aligned with the feature. When misaligned by even one data-point, however, the positive correlation can be negligible, and immediately another \xk (\textit{e.g.}, a null template) or even no particular \xk can dominate the BMS calculation. The effect of this behavior is that only the location in \y corresponding to the center of a feature of interest (\textit{i.e.}, perfectly overlapped by \xk) is identified with a high $P(\xk\vert\y,M_0)$. Interestingly, this means BITS inherently protects against multiple localisations of the same feature while simultaneously facilitating localisation of multiple, closely spaced features.
\\
\indent \indent 
Perhaps the most important and unique aspect of the BITS algorithm is that it enables \ys acquired under different experimental conditions to be directly compared within a single, common reference frame. Specifically, since BITS uses evidence expressions that are based only upon agreement of an \xk with the latent structure of \y, it is independent of many experimental nuisances that would otherwise obstruct direct comparisons. For example, while different background levels in two experiments can render the use of a common threshold value to localise features completely ineffective, BITS remains invariant under changes in background and thus yields two sets of feature localisations that can be directly compared. Moreover, by setting a pre-defined posterior probability threshold and/or by including a model of the background or simply noise as \xks in \xset, BITS can automatically identify and localise features for a wide range of experimental techniques without human intervention or advanced knowledge about the experimental situation (\textit{e.g.}, exposure times for a detector). 

\section{Discussion} \label{examples}
\begin{figure}
\centering
\vspace{-5mm}
\includegraphics[width=0.75\columnwidth]{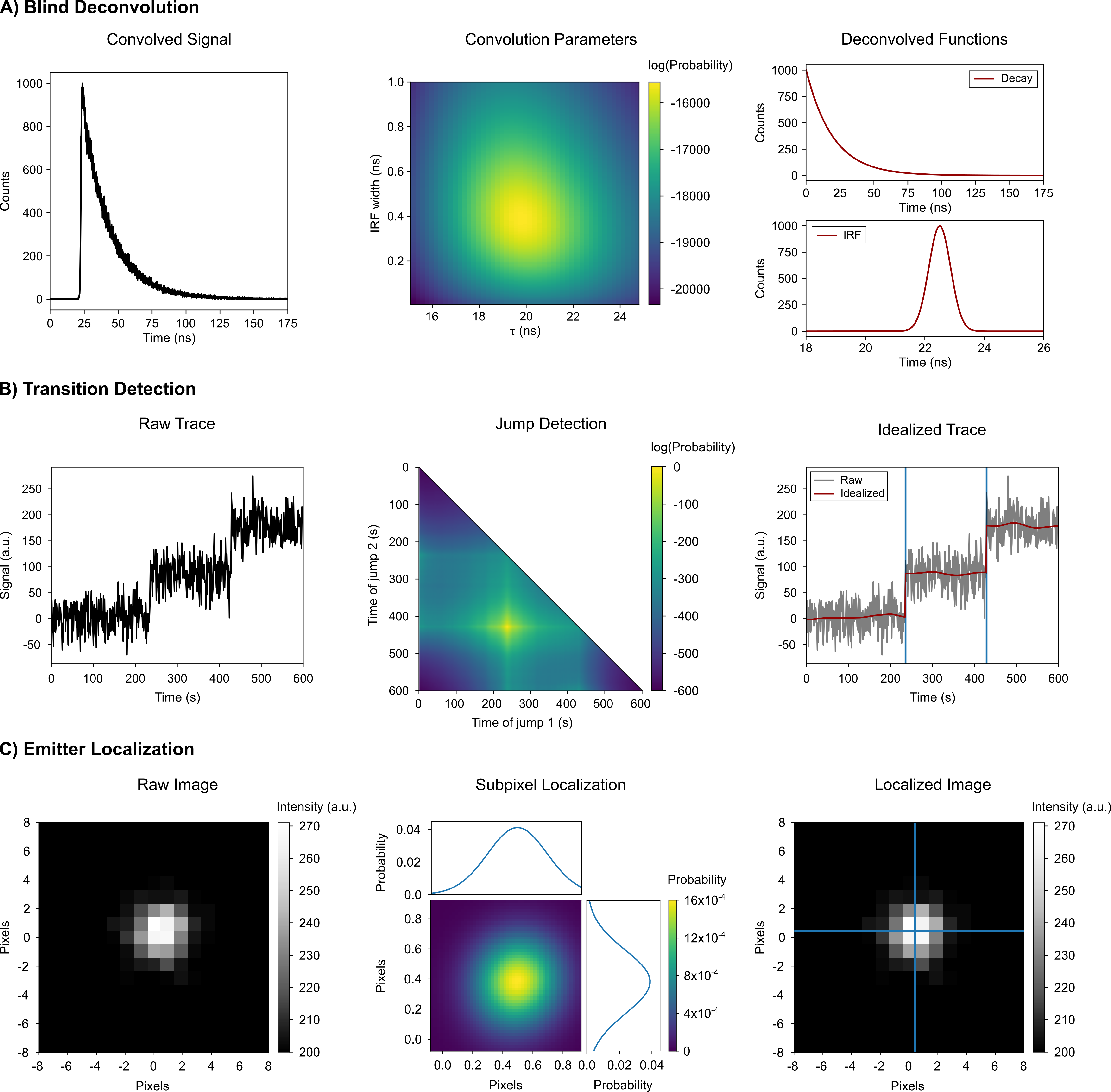}
\caption{\textbf{Examples of analyses based on shape calculations.} A) A fluorescence lifetime dataset (left) may be modeled as a convolution of an exponential decay of unknown lifetime ($\tau$) and a Gaussian instrument response function (IRF) of unknown width. By comparing shapes of the data with convolved templates, a joint log-probability map of the two unknowns is constructed (middle) and the deconvolved functions corresponding to the parameters with the maximum probabilities are plotted (right). B) A signal vs. time trajectory (left) with two discontinuous jumps may be compared in shape to a set of templates corresponding to all possible jump times to generate a map of the joint log-probabilities for the times of the jumps (middle).  The times corresponding to the maximum probabilities are overlaid (in blue) over the raw signal and the continuous segments identified are idealised (in red) using a Gaussian filter (right). C) A fluorescence emitter in a microscope image (left) may be modeled as a Gaussian of known width centered at a certain location. Using a subpixel grid to generate such templates with varying centers, a map of the joint probability for the co-ordinates of the emitter is plotted, along with marginalised probabilities for the x- and y-axes (middle). The co-ordinates with the maximum probability are overlaid (in blue) over the raw image (right).}
\vspace{-4mm}
\label{fig3} 
\end{figure}

\noindent
The ability to describe the latent structure of experimental data can readily be leveraged for many different techniques and analyses in the physical and life sciences (Fig. \ref{fig1}). Here, we consider several examples, and briefly demonstrate and discuss some of their implementations. We begin with experiments in which raw data is pre-processed in a manner according to its shape. For instance, in signal processing-based  analyses, the shape-based framework presented here can be used to: (i) perform multivariate calibration transfer \cite{Workman2018} (\textit{e.g.}, in optical spectroscopy techniques) by comparing the shapes of responses for standardised samples across multiple instruments; (ii) perform blind deconvolution \cite{Levin2009} (\textit{e.g.}, in atomic force-, optical-, or electron microscopy) by comparing the convolution of possible instrument response functions and deconvolved signals to the shape of the recorded data (Fig. \ref{fig3}A) ; or (iii) to align distinct measurements of the same object or sample (\textit{e.g.}, the same field-of-view in different colour channels \cite{Friedman2015} or multiple image planes \cite{Juette2008} of a fluorescence microscope) by finding the optimal polynomial transform to create an interpolated measurement that matches the shape of another measurement. 

\indent \indent 
In analytical chemistry experiments that are used to identify the contents of an experimental sample based on the characteristics, or ‘fingerprints’, of known standards, the framework presented here can be used to compare the shape of the signal of an experimental sample against databases of standard experimental and/or theoretical templates of the possible contents. Examples of this kind of analysis include identification of chemicals from characteristic infrared (IR) spectra \cite{Luinge1990}, or \textsuperscript{1}H nuclear magnetic resonance (NMR) spectra \cite{Napolitano2013}, and identification of proteins from fragmentation patterns in mass spectrometry data \cite{Domon2006}. Similarly, a shape assignment-based approach would enable the automated identification of sample contents from the noisy measurements typically obtained in analyses involving small sample concentrations. Additionally, shape comparisons can be used to construct and validate atomic- or near-atomic resolution models in structural biology experiments by comparing the: (i) structure factors from different molecular models to the diffraction pattern for an X-ray crystallography experiment \cite{Brunger1992}, (ii) electrostatic potential maps from different molecular models to the reconstructed density in a cryogenic electron microscopy (cryo-EM) experiment \cite{Scheres2012}, or even (iii) predicted electron scattering patterns from different atomic models to the raw electron microscopy micrographs in a cryo-EM experiment. In fact, in the last example, a similar Bayesian approach has been pioneered by Cossio and Hummer to analyse cryo-EM micrographs \cite{Cossio2013}. Despite differences in the priors, scaling parameter, and specific use of BMS, our generalised shape-based analysis framework otherwise readily maps onto this approach, and could thus be used
to develop a specialised method that achieves effectively the same results.
\\
\indent \indent 
In analyses in which transitions between different signal values need to be identified, our approach can be extended to locate change points by comparing \xks with and without a discrete change of any arbitrary magnitude in shape (Fig. \ref{fig3}B). This could be used to locate changes in the: (i) efficiency of fluorescence resonance energy transfer (E\textsubscript{FRET}) in the E\textsubscript{FRET} vs. time trajectories reporting on the (un)folding or conformational dynamics of a biomolecule in single-molecule FRET (smFRET) experiments \cite{Tinoco2011}, (ii) extension in the force vs. extension or extension vs. time trajectories reporting on the (un)folding or conformational dynamics of a biomolecule in single-molecule force spectroscopy experiments \cite{Tinoco2011, Bustamante2021}, (iii) position in the position vs. time trajectories reporting on the directional stepping of a biomolecular motor in single-molecule fluorescence experiments \cite{Park2007}, (iv) conductance in the conductance vs. time trajectories reporting on the (un)folding or conformational dynamics of a biomolecule in single-molecule field effect transistor (smFET) experiments \cite{Bouilly2016}, or even (v) conformation in the conformation vs. time trajectories reporting on (un)folding or conformational dynamics of a biomolecule in molecular dynamics simulations. Indeed, such a BMS-based method to detect transition in time-series data has been pioneered by Ensign and Pande \cite{Ensign2010}. Despite minimal differences in noise models, our generalised shape-based analysis framework can be mapped onto the  approach of Ensign and Pande, thus facilitating the development of a specialised method that can effectively arrive at the same results. This general approach to analysing time-series extends the use of Bayesian inference-based techniques for the detection of change points in a time-dependent signal to any sequential data with arbitrary signal properties, thereby enabling the accurate estimation of kinetics from a wide range of experimental techniques. 

\begin{figure}[ht]
\centering
\vspace{5mm}
\includegraphics[width=\columnwidth]{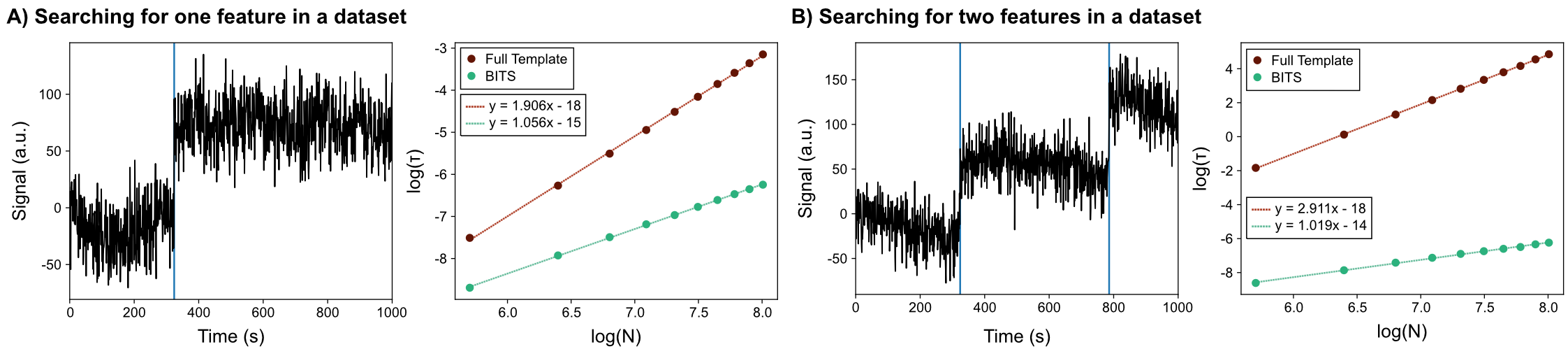}
\caption{\textbf{Computational efficiency of Full-dataset \textit{versus} BITS Analyses.} A) An example signal \textit{versus} time trajectory (left) with one discontinuous jump analysed by comparing the shape of the data to templates the size of the entire dataset (Full Template) and using BITS with a small template encoding a `step up'. The logarithm of the computational time, required to perform an analysis, $\tau$, is shown as a function of the logarithm of the length of the signal \textit{versus} time trajectory, $N$. The linearised curves were fit with a first-order polynomial to yield the computational scaling of each calculation. These values match the predicted scaling for one template ($R=1$) of $\mathcal{O}(N^{R+1}=N^2)$ and $\mathcal{O}(N^1)$, for Full Template and BITS, respectively. B) As in panel A, but with signal \textit{versus} time trajectories (left) with two discontinuous jumps such as in Fig. 3B. The computational scaling matches the predicted scaling for two templates ($R=2$) of $\mathcal{O}(N^3)$ and $\mathcal{O}(N^1)$. Together these results demonstrate the linear scaling of BITS with respect to the number of features in a dataset.}
\label{fig4}
\end{figure}

\indent \indent 
In addition to the full shape-based methods discussed above, we expect that BITS is poised to have a large impact on a number of more specialised situations--particularly analyses that require localisation of well-defined signals or image features. For example, given a particular spectral line shape (\textit{e.g.}, a Lorentzian function), BITS can be used to find peaks in multidimensional NMR spectra \cite{Hiller2005}. Similarly, BITS can be used to identify particles of interest and their orientations in cryo-EM \cite{FrankEM2006} and cryogenic electron tomography (cryo-ET) \cite{FrankET2006} micrographs in a manner similar to that shown in Fig. \ref{fig2}. This is also readily extended to localising and identifying individual molecules and molecular structures in atomic force microscopy (AFM) \cite{Dufrene2017} and super-resolution fluorescence microscopy images \cite{Khater2020} (Fig. \ref{fig3}C). Methods using traditional template matching for such analysis tasks can be easily adapted into our BITS framework, thereby enabling comparison across disparate datasets (\textit{e.g.}, \cite{amyot2020}, \cite{shi2019}). In the case of the analysis of time-series (\textit{e.g.}, single-molecule fluorescence, molecular dynamics), BITS can be used to detect transitions or change points between states, even when those transitions are more diffusive than instantaneous, as is a typical requirement for analysis using hidden Markov models. Additionally, as discussed in Sec. \ref{BITS}, BITS makes this type of time-series analysis much more efficient than when analysing the shape of a whole time-series at once (Fig. \ref{fig4}).
\\
\indent \indent 
The range of examples provided above are broad, but not exhaustive. Nonetheless, they highlight the versatility of our approach, and we hope they will inspire others to adopt this framework for their experiments and analysis methods. Although the development, optimisation, bench-marking, and in-depth discussion of each of the individual scientific applications described above is necessarily very specialised, and, thus, beyond the scope of the current work, we have created a gallery of illustrative, proof-of-principle examples that are open-source and written in Python to demonstrate and enable the use of our framework; they can be accessed at \url{https://bayes-shape-calc.github.io}.

\section{Conclusion} \label{conclusion}
\noindent
To the best of our knowledge, the use of probabilities to determine the latent structure of data as discussed above is a radically new approach to analysing experiments in the physical and life sciences. The framework we present here is the quantitative extension of a very intuitive approach to data analysis in which expert researchers visually determine whether their data is the shape that they expect it to be. Rather than develop heuristic approaches to emulate this subjective process, our method provides a quantitative metric based only on probability that is free from human intervention and experimental considerations. Among other things, the ability to determine the shape of data enables researchers to objectively pre-process data; identify fingerprints and validate assignments; detect change points; and identify and localise features using BITS (Figs. \ref{fig1} and \ref{fig2}).  Additionally, the shape-based framework we present here can be readily applied to analyse large, high-dimensional datasets that are difficult to visualise and would be nearly impossible to analyse manually. As can be seen from the breadth of potential applications listed above (Fig. \ref{fig1}), the overall methodology described in this work transcends individual fields and techniques, and indeed represents a new quantitative lens through which countless experiments in many different areas of the physical and life sciences may be analysed.

\begin{acknowledgments}

\noindent
This work was supported by funds to R.L.G. from the National Institutes of Health (NIH) (R01 GM 084288, R01 GM 137608, R01 GM 128239, and R01 GM 136960) and the National Science Foundation (NSF) (CHE 2004016) as well as funds to C.D.K. from the NIH (Training Grant in Molecular Biophysics to Columbia University, T32 GM008281), the Department of Energy (DOE) (Office of Science Graduate Fellowship, DE-AC05-06OR23100), and the NSF (CHE 2137630).
\end{acknowledgments}

\nocite{*}

\bibliography{BITS_MS_ArXiv}

\end{document}


\thispagestyle{plain}
\noindent
{\Large{\textbf{Supplemental Materials}}}
\vspace{\baselineskip}
\vspace{\baselineskip}

\noindent
{\textbf{\Large{Inferring the shape of data: A probabilistic framework for analysing experiments in the natural sciences}}}
\vspace{\baselineskip}

\noindent
{Korak Kumar Ray$^{1,*}$, Anjali R. Verma$^{1,*}$, Ruben L. Gonzalez, Jr.$^{1,\dagger}$, and Colin D. Kinz-Thompson$^{2,\dagger}$}
\vspace{\baselineskip}

\noindent
${}^1$\textit{Department of Chemistry, Columbia University, New York, NY 10027 USA}\\
${}^2$\textit{Department of Chemistry, Rutgers University-Newark, Newark, NJ 07102 USA}
\vspace{\baselineskip}

\noindent
${}^*$These authors contributed equally to this work.
\vspace{\baselineskip}

\noindent
${}^\dagger$To whom correspondence may be addressed: 
Ruben L. Gonzalez, Jr, Department of Chemistry, Columbia University, 3000 Broadway, New York, NY 10027, Tel: (212) 854-1096, Email: rlg2118@columbia.edu; Colin D. Kinz-Thompson, Department of Chemistry, Rutgers University-Newark, 73 Warren St, Newark, NJ 07102, Tel: (973) 353-0671, Email: colin.kinzthompson@rutgers.edu.

\section{Maximum entropy prior distributions}
\noindent
The maximum entropy approach relates informational constraints to specific probability distributions, so it is well suited to deriving the mathematical form of prior probability distributions\cite{jaynes}. Specifically, the maximum entropy prior for any given constraint is the probability distribution which has the maximum information entropy subject to those constraints. For an arbitrary continuous probability distribution $p(\theta)$, the entropy is
\begin{equation*}
    H = - \int p(\theta\vert I) \ln \left(\frac{p(\theta\vert I)}{m(\theta)}\right) d\theta,
\end{equation*}
where $\theta$ is a model parameter, $I$ represents our background conditional information, and $m(\theta)$ is the invariant measure\cite{jaynes}. Given any constraint on $\theta$, the above term for entropy may be maximised using the method of Lagrange multipliers to obtain the corresponding maximum entropy prior distribution. All probability distributions are subject to a normalisation constraint, such that
\begin{equation*}
    \int_{-\infty}^{\infty} p(\theta) d\theta = 1.
\end{equation*}

\indent \indent
In this section, we derive several maximum entropy priors that are utilised in the subsequent sections and in the main text of the manuscript. It should be noted that these only form a subset of the possible maximum entropy priors, which can be generated for any given number of constraints.

\subsection{Uniform prior} \label{ssec:uniform}
\noindent
If the only prior information for a parameter $\theta$ is that $\theta_{\text{min}} \le \theta \le \theta_{\text{max}}$ and there is no reason to favour one value of $\theta$ within this range over another, then the invariant measure is $m(\theta)=1$, and the corresponding Lagrange function for the prior probability distribution can be written using $\lambda$ as a Lagrange multiplier as
\begin{equation*}
    Q =  - \int p(\theta\vert I) \ln (p(\theta\vert I)) d\theta + \lambda \left( 1-\int p(\theta\vert I) d\theta \right),
\end{equation*}
which, when maximised, yields
\begin{equation*}
    \lambda = - \ln p(\theta\vert I).
\end{equation*}
Ensuring the prior is normalised then yields 
\begin{equation*}
    p(\theta\vert I) = \frac{1}{\theta_{\text{max}} - \theta_{\text{min}}} \equiv \frac{1}{\Delta \theta} \quad \text{for } \theta_{\text{min}} \le \theta \le \theta_{\text{max}},
\end{equation*}
which is the continuous uniform distribution.

\subsection{Log-uniform prior}
\noindent
A similar situation to that in Section \ref{ssec:uniform} is when $\theta_{\text{min}} \le \theta \le \theta_{\text{max}}$ but there is no reason to favour one particular magnitude of $\theta$ over another (\textit{i.e.}, $\ln \theta$). In this case, the invariant measure is $m(\theta)=\theta^{-1}$, and the corresponding maximum entropy prior is 
\begin{equation*}
    p(\theta\vert I) = \frac{\theta^{-1}}{\ln\theta_{\text{max}} - \ln\theta_{\text{min}}} \equiv \frac{\theta^{-1}}{\Delta \ln \theta} \quad \text{for } \theta_{\text{min}} \le \theta \le \theta_{\text{max}}.
\end{equation*}

\indent \indent
In this situation, the prior is uniform on a logarithmic scale, and is useful for scale parameters such as variance or precision.

\subsection{Gaussian prior}
\noindent
For a situation where the expectation value and the variance of the parameter $\theta$ are known to be $\mu$ and $\tau^{-1}$, respectively, we have the constraints that
\begin{align*}
    \E[\theta] &= \int_{-\infty}^{\infty} \theta p(\theta\vert I) d\theta \equiv \mu, \text{ and }\\
    \E[\theta^2 - \E[\theta]^2] &= \E[\theta^2] - \mu^2 = \int_{-\infty}^{\infty} \theta^2 p(\theta\vert I) d\theta - \mu^2 \equiv \frac{1}{\tau}.
\end{align*}
To derive the maximum entropy prior for this situation, when the invariant measure $m(\theta)=1$, we write the Lagrange function with Lagrange multipliers $\alpha$, $\beta$, and $\gamma$ as
\begin{align*}
    Q = - \int_{-\infty}^{\infty} p(\theta\vert I) \ln (p(\theta\vert I)) d\theta + \alpha \left(1-\int_{-\infty}^{\infty} p(\theta\vert I) d\theta\right) &+ \beta \left(\mu-\int_{-\infty}^{\infty} \theta p(\theta\vert I) d\theta \right) \\
    &+\gamma \left( (\mu^2+\tau^{-1}) - \int_{-\infty}^{\infty} \theta^2 p(\theta\vert I) d\theta \right).
\end{align*}
When maximised, this yields
\begin{equation*}
    p(\theta\vert I) = \frac{\tau}{\sqrt{2\pi}}e^{-\frac{\tau}{2}(\theta - \mu)^2},
\end{equation*}
which is a Normal or Gaussian distribution. Thus, the use of a Gaussian prior represents previous information on the mean and the variance of the parameter.

\subsection{Gamma prior}
\noindent
For a situation where the expectations of the parameter and the logarithm of the parameter are known to be $\mu$ and $L$, respectively, we have the constraints that
\begin{align*}
    \E[\theta] &=  \int \theta p(\theta\vert I) d\theta = \mu , \\
    \E[\ln \theta] &=  \int \ln \theta p(\theta\vert I) d\theta = L,
\end{align*}
and, given the logarithmic function, that $\theta > 0$. To derive the maximum entropy prior for this situation, we write the Lagrange function with Lagrange multipliers $\alpha$, $\beta$, and $\gamma$ as
\begin{align*}
    Q = - \int_{0}^{\infty} p(\theta\vert I) \ln (p(\theta\vert I)) d\theta + \alpha \left(1-\int_{0}^{\infty} p(\theta\vert I) d\theta\right) &+ \beta\left( \mu - \int_{0}^{\infty} \theta p(\theta\vert I) d\theta\right) \\
    &+ \gamma \left(L - \int_{0}^{\infty} \ln \theta p(\theta\vert I) d\theta \right).
\end{align*}
Upon maximisation, this yields the prior 
\begin{equation*}
    p(\theta\vert I) = \frac{\beta^\alpha}{\Gamma(\alpha)}\theta^{\alpha - 1}e^{-\beta \theta},
\end{equation*}
where $\Gamma(\alpha)$ denotes the gamma function. This probability distribution is known as the gamma distribution.

\subsection{A joint prior for modeling shapes}
\noindent
In the background information used in this work to calculate the shape of data, we have assumed that the three parameters in the main text of the manuscript ($m$, $b$, and $\tau$) are independent of each other. This assumption yields the joint prior probability distribution
\begin{equation*}
    p(m, b, \tau| M_0) = p(m| M_0) p(b| M_0) p(\tau| M_0),
\end{equation*}
where the conditional $M_0$ represents our background knowledge for the inference problem. 

\indent \indent
For $m$, we make assumptions that lead us to use either a uniform, Gaussian, or gamma maximum entropy prior, which are
\begin{align*}
       p(m| M_0) &= \frac{1}{\Delta m},  \\
       p(m| M_0) &= \frac{\Lambda_m}{\sqrt{2\pi}} e^{-\frac{\Lambda_m}{2}(m - \mu_m)^2}, \qquad \text{ or} \\
       p(m| M_0) &= \frac{{\beta_m}^{\alpha_m}}{\Gamma(\alpha_m)}m^{\alpha_m - 1}e^{-\beta_m m}.
\end{align*}
\indent \indent
For $b$, we have only made assumptions the lead us to use a uniform maximum entropy prior, which is
\begin{equation*}
    p(b| M_0) = \frac{1}{\Delta b}.
\end{equation*}
\indent \indent
For $\tau$, we have made the assumption that we do not know the magnitude of the noise and thus use a log-uniform maximum entropy prior, which is 
\begin{equation*}
    p(\tau| M_0) = \frac{\tau^{-1}}{\Delta \ln \tau}.
\end{equation*}

\section{Evidence expressions for calculating the shape of data}
\subsection{Definitions}
\noindent
In the derivations presented in this section, we use the following terms and definitions:
\begin{align*}
    N &\qquad: \text{Number of data points} \\
    M &\qquad\equiv \frac{N}{2}-1 \\
    \braket{\x} &\qquad: \text{Average of }\x \biggl(\equiv \frac{1}{N}\sum_{i=1}^N x_i\biggr) \\
    V_x &\qquad\equiv \braket{\x^2} - \braket{\x}^2 \\
    V_y &\qquad\equiv \braket{\y^2} - \braket{\y}^2 \\
    C &\qquad\equiv \braket{\x\y} - \braket{\x}\braket{\y} \\
    r &\qquad\equiv \frac{C}{\sqrt{V_x V_y}} \\
    \text{erf}(\alpha) &\qquad: \text{Error function} \\
    B(\alpha, \beta) &\qquad: \text{Beta function} \\
    I_{\nu}(\alpha,\beta) &\qquad: \text{Regularised incomplete Beta function} \\
    \pFq{2}{1}{\alpha,\beta}{\gamma}{z} &\qquad: \text{Gaussian hypergeometric function}
\end{align*}

\subsection{Gaussian likelihood}\label{ssec:gaussian}
\noindent
The evidence functions in the following sub-sections are all computed assuming in $M_0$ that the likelihood function is a Gaussian distribution (\textit{i.e.}, white noise). Specifically, this likelihood function for the data, $\y$, with the template, $\x$, is
\begin{equation*}
    p(\y | \x, m, b, \tau, M_0) = \prod_{i=1}^{N} \frac{\tau}{\sqrt{2\pi}} e^{-\frac{\tau}{2}(y_i - m x_i - b)^2}.
\end{equation*}
The integrals are computed using standard tabulated forms.\cite{gradshteyn,ng}

\subsubsection{Uniform priors for $\mathbf{m}$ and $\mathbf{b}$, log-uniform prior for $\mathbf{\tau}$; $\mathbf{m \in \R, b \in \R, \tau > 0}$} 
\noindent
Intermediate integration steps are shown for this section. All other cases are computed using the same order of integration (\textit{i.e.}, first $b$, then $m$, then $\tau$).

\begin{align*}
    p(\y|\x,m,\tau, M_0)&=\int_{-\infty}^{\infty} db \; p(\y | \x, m, b, \tau, M_0) p(b| M_0) \\
                         &= \int_{-\infty}^{\infty} db \frac{\tau^{N/2}}{(2\pi)^{N/2} \Delta b}  e^{-\sum_{i=1}^N \frac{(y_i - m x_i - b)^2 \tau}{2}} \\
                         &= \frac{\tau^{(N-1)/2}}{(2\pi)^{(N-1)/2} \Delta b \sqrt{N}} e^{- \frac{N\tau}{2}(V_x + m^2 V_y - 2m C)}
\end{align*}

\begin{align*}
    p(\y|\x, \tau, M_0)&=\int_{-\infty}^{\infty} dm \; p(\y | \x, m, \tau, M_0) p(m| M_0) \\
                        &= \int_{-\infty}^{\infty} dm  \frac{\tau^{(N-1)/2}}{(2\pi)^{(N-1)/2} \Delta b \Delta m \sqrt{N}} e^{- \frac{N\tau}{2}(V_x + m^2 V_y - 2m C)} \\
                        &= \frac{\tau^{M}}{N(2\pi)^{M} \Delta b \Delta m  \sqrt{V_x}} e^{- \frac{N\tau}{2}V_y (1 - r^2)} 
\end{align*}

\begin{align*}
    P(\y | \x, M_0) &= \int_{0}^{\infty} d\tau \; p(\y | \x, \tau, M_0) p(\tau| M_0) \\
                       &= \int_{0}^{\infty} d\tau \frac{\tau^{M-1}}{N(2\pi)^{M} \Delta b \Delta m \Delta \ln \tau \sqrt{V_x}} e^{- \frac{N\tau}{2}V_y (1 - r^2)}  \\
                       &= \frac{\Gamma(M)N^{-\frac{N}{2}} V_x^{-\frac{1}{2}}}{\Delta b \Delta m \Delta \ln \tau} [\pi V_y(1-r^2)]^{-M}.
\end{align*}

\subsubsection{Uniform priors for $\mathbf{m}$ and $\mathbf{b}$, log-uniform prior for $\mathbf{\tau}$; $\mathbf{m > 0, b \in \R, \tau > 0}$} \label{sssec:bitsintegral}
\noindent
This integral is used in the main text of the manuscript and in Section \ref{sec:modelselection}. From above,
\begin{align*}
    p(\y|\x,m,\tau, M_0) &= \frac{\tau^{(N-1)/2}}{(2\pi)^{(N-1)/2} \Delta b \sqrt{N}} e^{- \frac{N\tau}{2}(V_x + m^2 V_y - 2m C)}
\end{align*}
Further,
\begin{align*}
    p(\y| \x, \tau) &= \int_{0}^{\infty} dm \; p(\y | \x, m, \tau, M_0) p(m| M_0) \\
    &=  \frac{\tau^{(N-1)/2 }}{(2\pi)^{M} \Delta b \Delta m N \sqrt{V_x}} e^{- \frac{N\tau}{2}V_y(1 - r^2)} \times \frac{1}{2} \biggl[ 1 + \text{erf}\biggl(\frac{C\sqrt{N \tau}}{ \sqrt{2 V_x}}\biggr) \biggr]
\end{align*}
\begin{align*}
    P(\y|\x, M_0)  &= \int_{0}^{\infty} d\tau \; p(\y | \x, \tau, M_0) p(\tau| M_0) \\
    &= \frac{\Gamma(M)N^{-\frac{N}{2}} V_x^{-\frac{1}{2}}}{\Delta m \Delta b \Delta \ln\tau} \left[\pi V_y(1-r^2)\right]^{-M} \times \\
    &\qquad \qquad \Bigg[\frac{1}{2} + r (1-r^2)^{-1/2} B\left(\frac{1}{2},M\right) \pFq{2}{1}{\frac{1}{2},M+\frac{1}{2}}{\frac{3}{2}}{\frac{-r^2}{1-r^2}} \Bigg]\\
    &= \frac{\Gamma(M) N^{-\frac{N}{2}} V_x^{-\frac{1}{2}} }{2\Delta m \Delta b \Delta \ln\tau} \left[\pi V_y(1-r^2)\right]^{-M} \left[ 1 + \frac{r}{|r|} I_{r^2}\left(\frac{1}{2},M\right) \right].
\end{align*}

\subsubsection{Uniform priors for $\mathbf{m}$ and $\mathbf{b}$, log-uniform for $\mathbf{\tau}$; $\mathbf{m>0, b>0, \tau > 0}$} \label{sssec:hard}
\noindent
While this condition is the most `realistic' model for some applications, a closed form for this integral,  to the best of our knowledge, unfortunately does not exist. Using different priors did not yield a closed form either. If required, this integral may be numerically calculated (\textit{e.g.}, using a triple quadrature method), however, this is much more computationally expensive than using an analytical form.

\begin{equation*}
    P(\y|\x, M_0) = \int_0^\infty \int_0^\infty \int_0^\infty d\tau \ dm \ db \ \frac{\tau^{-1}}{\Delta b \Delta m \Delta \ln\tau} \left(\frac{\tau}{2\pi}\right)^{\frac{N}{2}}e^{-\frac{\tau}{2}\sum_{i=1}^N (y_i - mx_i - b)^2}.
\end{equation*}

\subsubsection{Uniform prior for $\mathbf{b}$, log-uniform prior for $\mathbf{\tau}$; $\mathbf{m = 0, b \in \R, \tau > 0}$} \label{sssec:mnull}
\noindent
This evidence function is for the case where there is no template, $\x$, since the condition $m = 0$ removes any dependence on $\x$. Thus, this evidence represents a `flat' shape composed only of background and noise. This integral is used in the main text of the manuscript and in Section \ref{sec:modelselection}.

\begin{equation*}
    P(\y| \x_{null}, M_0) = \frac{\Gamma (M +\frac{1}{2}) N^{-\frac{N}{2}}}{\Delta b \Delta \ln\tau} ( \pi  V_y)^{-(M + \frac{1}{2})}.
\end{equation*}

\subsubsection{Uniform prior for $\mathbf{b}$, log-uniform prior for $\mathbf{\tau}$; $\mathbf{m = 1, b \in \R, \tau > 0}$}

\begin{equation*}
    P(\y| \x, M_0) = \frac{\Gamma ( M + \frac{1}{2}) N^{-\frac{N}{2}}}{\Delta b \Delta \ln \tau} \left[ \pi \left( V_y + V_x - 2C\right)\right]^{-(M + \frac{1}{2})}.
\end{equation*}

\subsubsection{Uniform prior for $\mathbf{m}$, log-uniform prior for $\mathbf{\tau}$; $\mathbf{m \in \R, b = 0, \tau > 0}$}
\begin{equation*}
    P(\y|\x, M_0) = \frac{\Gamma ( M + \frac{1}{2}) N^{-\frac{N}{2}} \braket{x^2}^{-\frac{1}{2}}}{\Delta m \Delta \ln \tau} \left[ \pi \left( \braket{y^2} - \frac{\braket{xy}^2}{\braket{x^2}}\right)\right]^{-(M + \frac{1}{2})}.
\end{equation*}

\subsubsection{Log-uniform prior for $\mathbf{\tau}$; $\mathbf{m = 1, b = 0, \tau > 0}$}
\begin{equation*}
    P(\y|\x, M_0) = \frac{\Gamma(\frac{N}{2}) }{ \Delta \ln \tau } [\pi N ( \braket{y^2} + \braket{x^2} - 2\braket{xy})]^{-\frac{N}{2}}.
\end{equation*}

\subsubsection{Gaussian prior for $\mathbf{m}$, uniform for $\mathbf{b}$, log-uniform for $\mathbf{\tau}$; $\mathbf{m \in \R, b \in \R, \tau > 0}$} \label{sssec:approx1}
\noindent
To the best of our knowledge, the closed form of this integral only exists under the assumption  $N V_x \gg \Lambda_m$,  in which case, the approximate integral is given below. While computationally more expensive, it is recommended that the value of the integral is computed numerically instead.
\begin{align*}
    P(\y|\x, M_0) &= \int_0^\infty \int_{-\infty}^\infty \int_{-\infty}^\infty d\tau \ dm \ db \ \frac{\tau^{\frac{N}{2}-1} \Lambda_m e^{-\frac{\Lambda_m}{2}(m - \mu_m)^2}}{(\sqrt{2\pi})^{\frac{N+1}{2}}\Delta b \Delta \ln\tau} e^{-\frac{\tau}{2}\sum_{i=1}^N (y_i - mx_i - b)^2} \\
    &\approx \frac{\Gamma(M)N^{-\frac{N}{2}} }{\Delta b \Delta \ln \tau } \left(\frac{\Lambda_m}{2\pi V_x} \right)^{\frac{1}{2}} e^{- \frac{\Lambda_m}{2}\left(\mu_m - \frac{C}{V_x}\right)^2}  [\pi V_y(1-r^2)]^{-M}.
\end{align*}

\subsubsection{Gaussian prior for $\mathbf{m}$, log-uniform prior for $\mathbf{\tau}$; $\mathbf{m \in \R, b = 0, \tau > 0}$}
\noindent
To the best of our knowledge, the closed form of this integral only exists under the assumption  $N \braket{x^2} \gg \Lambda_m$,  in which case, the approximate integral is given below. While computationally more expensive, it is recommended that the value of the integral is computed numerically instead.
\begin{align*}
    P(\y|\x, M_0) &= \int_0^\infty \int_{-\infty}^\infty  d\tau \ dm \  \frac{\tau^{\frac{N}{2}-1} \Lambda_me^{-\frac{\Lambda_m}{2}(m - \mu_m)^2}}{(\sqrt{2\pi})^{\frac{N+1}{2}}\Delta b \Delta \ln\tau} e^{-\frac{\tau}{2}\sum_{i=1}^N (y_i - mx_i - b)^2} \\
    &\approx \frac{\Gamma(M + \frac{1}{2})N^{-\frac{N}{2}} }{\Delta \ln \tau } \left(\frac{\Lambda_m}{2\pi \braket{x^2}} \right)^{\frac{1}{2}} e^{- \frac{\Lambda_m}{2}\left(\mu_m - \frac{\braket{xy}}{\braket{x^2}}\right)^2}\left[ \pi \left( \braket{y^2} - \frac{\braket{xy}^2}{\braket{x^2}}\right)\right]^{-(M + \frac{1}{2})}.
\end{align*}

\subsubsection{Gamma prior for $\mathbf{m}$, uniform for $\mathbf{b}$, log-uniform for $\mathbf{\tau}$; $\mathbf{m > 0, b \in \R, \tau > 0}$} \label{sssec:approx2}
\noindent
To the best of our knowledge, the closed form of this integral only exists under the assumption  $N V_x \gg \beta_m$,  in which case, the approximate integral is given below. While computationally more expensive, it is recommended that the value of the integral is computed numerically instead.

\begin{align*}
    P(\y|\x, M_0) = \int_0^\infty \int_{0}^\infty \int_{-\infty}^\infty d\tau \ dm \ db \  &\frac{\tau^{\frac{N}{2}-1} {\beta_m}^{\alpha_m} m^{\alpha_m - 1}e^{-\beta_m m} }{(\sqrt{2\pi})^{\frac{N}{2}} \Gamma(\alpha_m)\Delta b \Delta \ln\tau} e^{-\frac{\tau}{2}\sum_{i=1}^N (y_i - mx_i - b)^2} \\ 
    \approx \frac{\beta_m^{\alpha_m} V_x^{-\frac{\alpha_m}{2}} V_y^{-\frac{N - \alpha_m - 1}{2}}N^{-\frac{N}{2}}}{(\pi)^{\frac{N-1}{2}}  \Gamma(\alpha_m) \Delta b  \Delta \ln \tau } &\Biggl\{ \frac{1}{2}\Gamma\biggl(\frac{N - (\alpha_m + 1)}{2}\biggr) \Gamma\biggl(\frac{\alpha_m}{2}\biggr) \pFq{2}{1}{\frac{\alpha_m}{2}, \frac{N - (\alpha_m +1)}{2}}{\frac{1}{2}}{r^2}  \\
    & + r \Gamma\biggl(\frac{N - \alpha_m }{2}\biggr) \Gamma\biggl(\frac{\alpha_m + 1}{2}\biggr) \pFq{2}{1}{\frac{\alpha_m + 1}{2}, \frac{N - \alpha_m}{2}}{\frac{3}{2}}{r^2} \Biggr\}.
\end{align*}

\subsection{Non-Gaussian Likelihoods}
\noindent
In Section \ref{ssec:gaussian}, we report evidence functions corresponding to models, $M_0$, where the likelihood is a Gaussian (\textit{i.e.}, for models with white noise). In other situations, the evidence for a model with a different likelihood  (\textit{e.g.}, $M'_0$) may be of interest. In particular, in the case of Poisson noise, the likelihood function would be
\begin{equation*}
    p(\y | \x, m, b, M'_0) = \prod_{i=1}^N \frac{(mx_i + b)^{y_i} e^{-(mx_i +b)}}{\Gamma(y_i)}.
\end{equation*}
As in Section \ref{sssec:hard}, the analytical form of the corresponding evidence function proved difficult to obtain by integrating the joint probability $P(\y,m,b\vert\x,M'_0)$. If required, it can be numerically integrated (\textit{e.g.}, using a double quadrature method).

\section{Simplifications for model selection} \label{sec:modelselection}
\noindent
The posterior probability for the $k^{\text{th}}$ template, $\xk$, given the dataset, $\y$, is
\begin{align*}
    P(\xk | \y, M_0) &= \frac{P(\y|\xk, M_0) P(\xk| M_0)}{\sum_{i=1}^K P(\y|\x_i, M_0) P(\x_i| M_0)} \\
    &= \frac{1}{\sum_{i=1}^K\frac{P(\y|\x_i, M_0) P(\x_i| M_0)}{P(\y|\xk, M_0) P(\xk| M_0)}} \\
    &\equiv \frac{1}{\sum_{i=1}^K R_{i,k} S_{i,k}},
\end{align*}
where $R_{i,k}$ denotes the ratio of the evidences for $\x_i$ over $\xk$, and $S_{i,k}$ denotes the ratio of the model priors for $\x_i$ over $\xk$. The former ratio, $R_{i,k}$, is known as a Bayes factor. In the following sections, we provide some simplified formulas for Bayes factors in which many terms have been canceled. We also note that it is possible that a significant amount of the approximation error in the approximate evidence functions given in Sections \ref{sssec:approx1} - \ref{sssec:approx2} cancel in such ratios.

\subsection{Identifying a shape from the background($\xk$ \textit{vs.} $\x_{null}$) }
\noindent
Consider calculating the shape of $\y$, where at least one model is given by the template $\xk$, and the likelihood and prior assumptions used in Section \ref{sssec:bitsintegral}, and where another of the models is the absence of a template (\textit{i.e.}, a flat background) as given by $\x_{null}$ in Section \ref{sssec:mnull}.  In this situation, calculating the probability of $\xk$ involves the Bayes factor
\begin{equation*}
    R_{\x_{null},\xk} =  \frac{2\Delta m}{B\left(\frac{1}{2},M\right)} \left( \frac{V_{{\xk}}}{V_{\y}} \right)^{\frac{1}{2}} \frac{\left( 1-r_k^2 \right)^M}{\left(1+\frac{r_k}{\lvert r_k \rvert} I_{r_k^2}(\frac{1}{2},M)\right)},
\end{equation*}
where subscript $k$ denotes terms related to $\xk$.

\subsection{Selecting between two templates ($\xk$ \textit{vs.} $\x_i$)}
\noindent
Consider calculating the shape of $\y$, when at least two of the models use templates, $\xk$ and $\x_i$, and both use the likelihood and prior assumptions from Section \ref{sssec:bitsintegral}. In this situation, calculating the probability of $\xk$ involves the Bayes factor
\begin{equation*}
    R_{\x_i,\xk} = \left(\frac{V_{\xk}}{V_{\x_i}}\right)^{1/2} \left(\frac{1-r_k^2}{1-r_i^2}\right)^{M} \frac{\left( 1+\frac{r_i}{\lvert r_i \rvert} I_{r_i^2}(\frac{1}{2},M) \right)}  {\left( 1+\frac{r_k}{\lvert r_k \rvert} I_{r_k^2}(\frac{1}{2},M) \right)},
\end{equation*}
where the subscripts denote the specified template.

\bibliography{BITS_SI_arxiv}